\documentclass[a4paper,11pt]{article}
\usepackage{jinstpub} 
\usepackage{lineno}

\title{\boldmath Optical surface reflectivity characterization in water Cherenkov detectors: methodologies and industrial applications}

\author{Deepak Tiwari}
\affiliation[1]{Institute for Basic Science,
55, Expo-ro, Yuseong-gu, Daejeon, Korea, 34126 \note{former affiliation}}

\emailAdd{deepaktiwari.ec@gmail.com}
\abstract{Understanding the optical properties of various components in water Cherenkov (WC) neutrino experiments is essential for accurate detector characterization, which is critical for precise measurements. Of particular importance is the characterization of surface reflectivity within the Cherenkov volume. We present a methodology for surface reflectivity characterization using a goniometer setup, addressing the challenges associated with measurements in the air and water (or other optical media). Additionally, we discuss the broader implications of Bidirectional Reflectance Distribution Function (BRDF) measurements using a goniometer, including their industrial applications.}
\begin{document}
\maketitle
\flushbottom
\section{Introduction}
\label{sec:intro}
\begin{figure}[htbp]
\centering
\includegraphics[width=.4\textwidth]{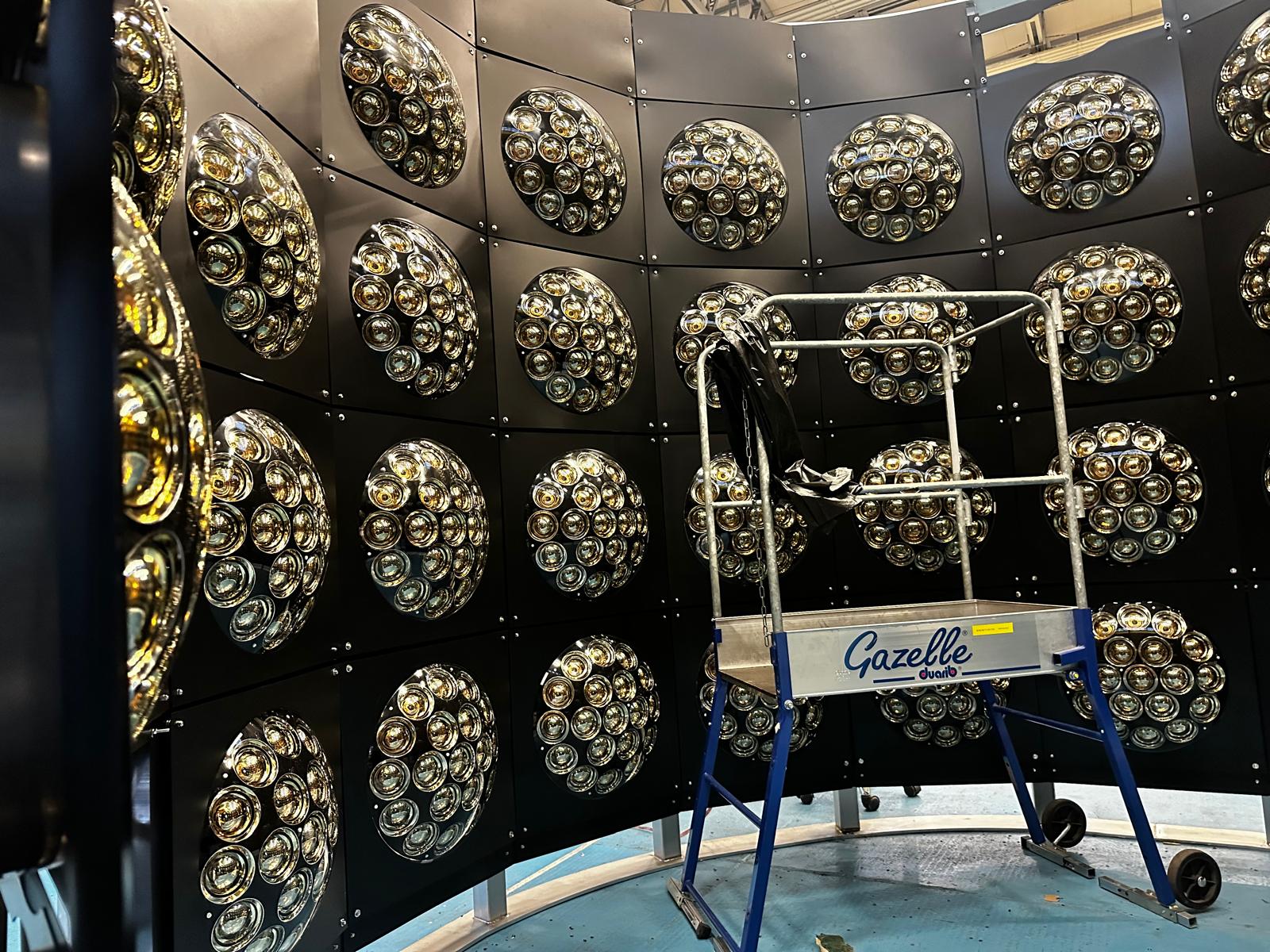}
\qquad
\includegraphics[width=.4\textwidth]{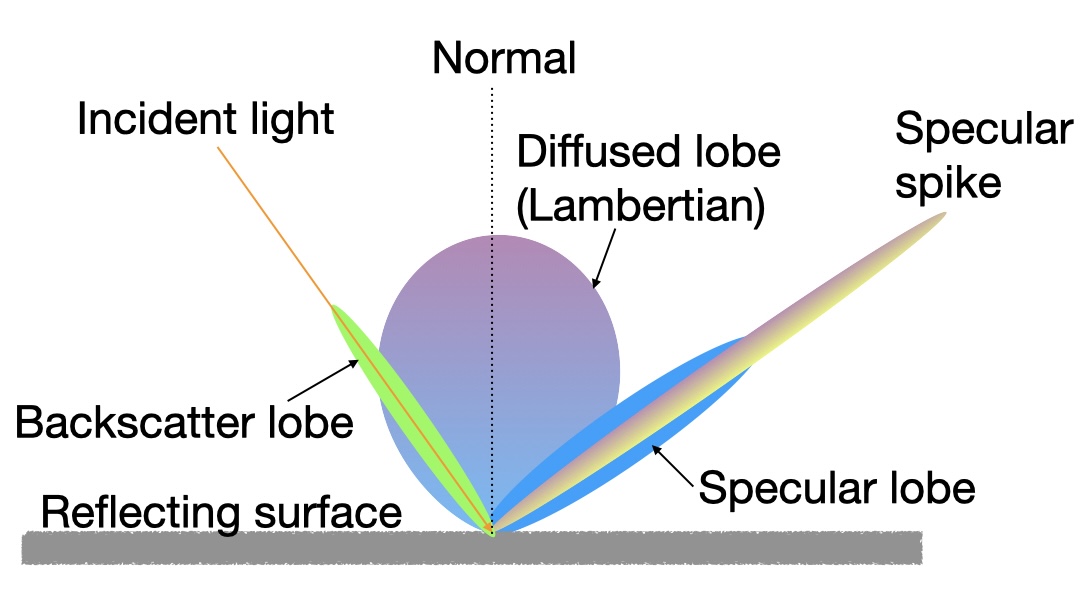}
\caption{Left: The black sheet, black Polyvinyl chloride (PVC), during the installation phase in the WCTE detector at CERN (courtesy: WCTE Collaboration). Right:  A schematic of light reflectance components, as included in the GEANT4 \cite{AGOSTINELLI2003250} UNIFIED Model for optical surfaces, comprising specular spike, specular lobe, diffuse (Lambertian) and backscatter. Please see ref.\cite{10.1117/1.JBO.19.2.026004} for further details.\label{fig:1}}
\end{figure}

The next generation of neutrino experiments\cite{Hyper-Kamiokande:2018ofw} geared towards precise neutrino measurements demands strong constraints on detector systematic errors.  In WC experiments, and particularly for the Water Cherenkov Test Experiment (WCTE)\cite{Barbi:2692463}, the `inner black lining' or `black sheet' of the detector surrounding the Cherenkov volume should have low total reflectivity. In addition, reflectivity should be ideally diffused without prominent specular components (see Figure~\ref{fig:1}). This is important to absorb stray photons and minimize the overall background, ensuring that only the direct Cherenkov light is detected by the photomultiplier tubes (PMTs), leading to a better signal-to-noise ratio. This is crucial to improve the precision of particle identification and energy reconstruction. With that goal, at the IBS, we performed detailed total reflectivity measurements of the candidate black sheet materials using an integrating sphere and angular component investigation using a goniometer.

\section{The methodology}
\label{sec:method}
\begin{figure}[htbp]
\centering
\includegraphics[width=.45\textwidth]{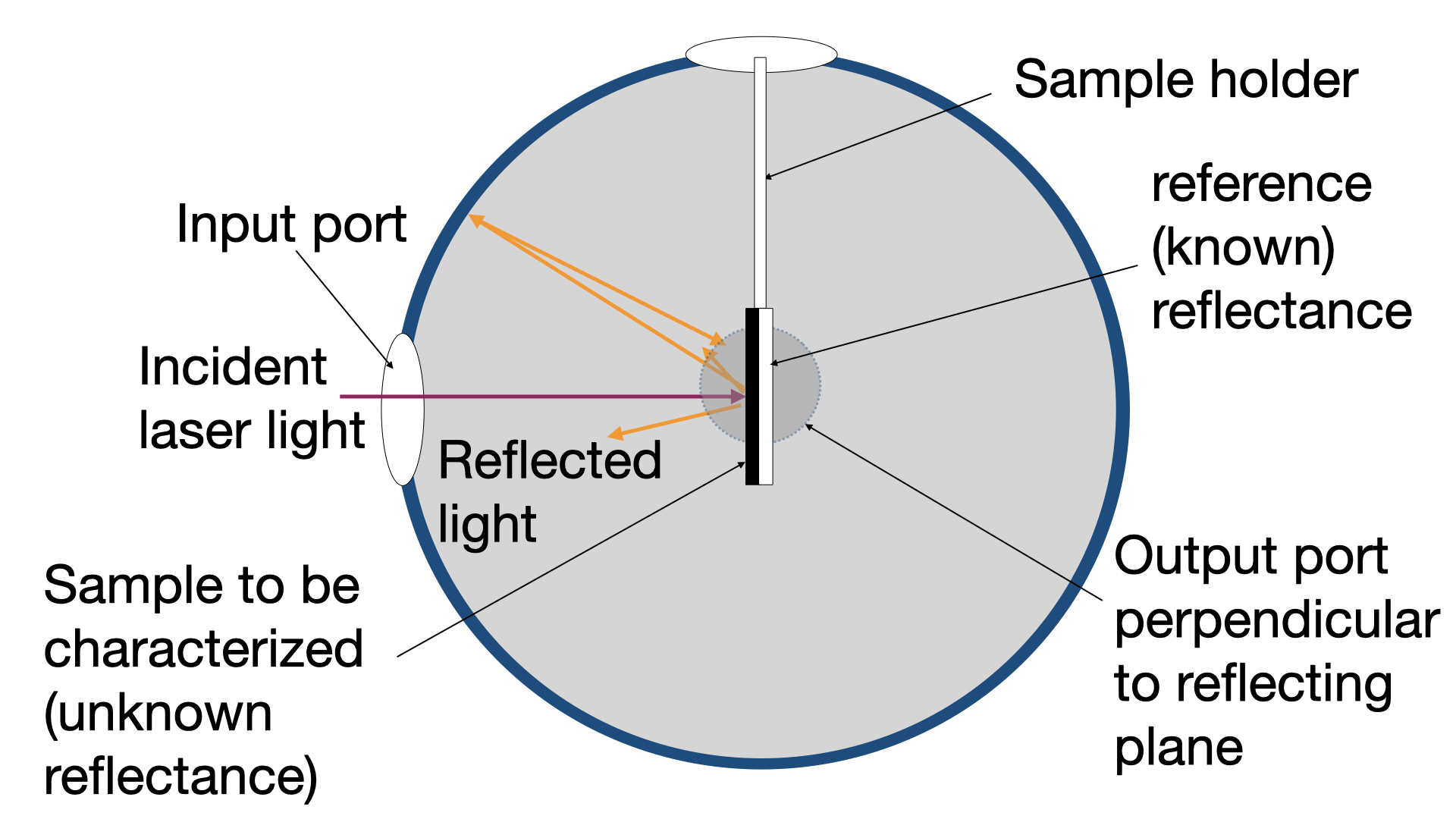}
\qquad
\includegraphics[width=.35\textwidth]{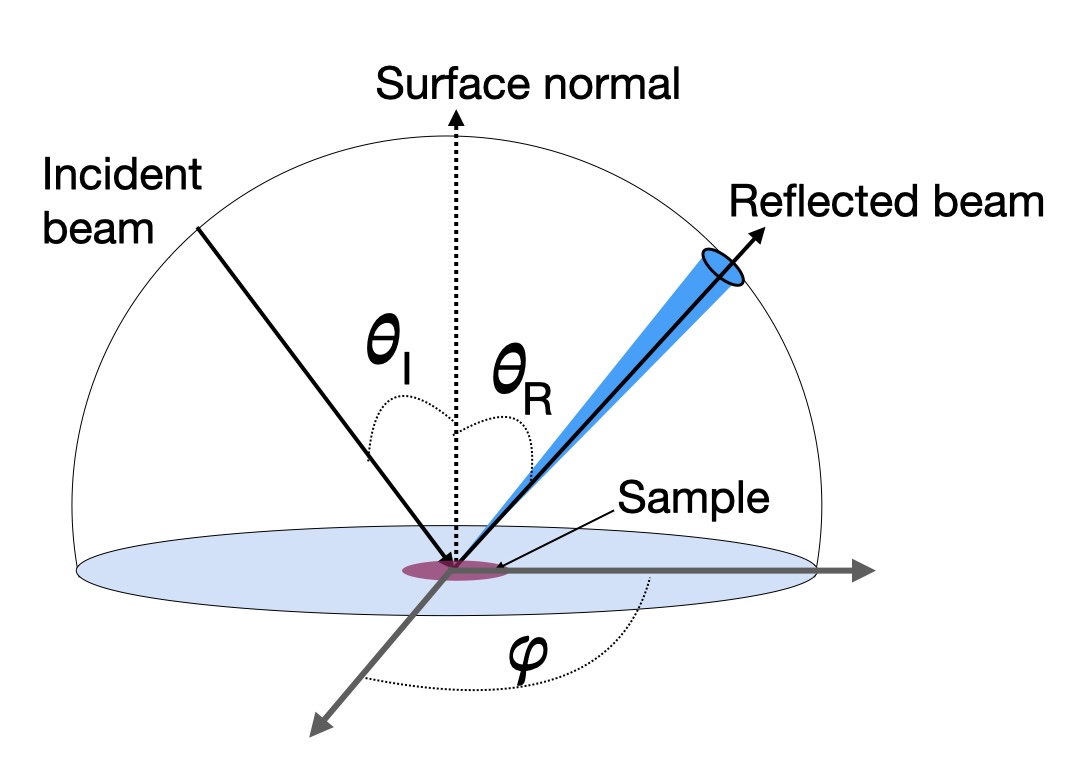}
\caption{Left: IS setup for total reflectivity measurements. Right: Goniometer schematic, $\theta$ and $\phi$ denote zenith and azimuth angles respectively, and $I$ and $R$ represents incident and reflected light, respectively.\label{fig:2}}
\end{figure}

\begin{figure}[htbp]
\centering
\includegraphics[width=0.85\textwidth]{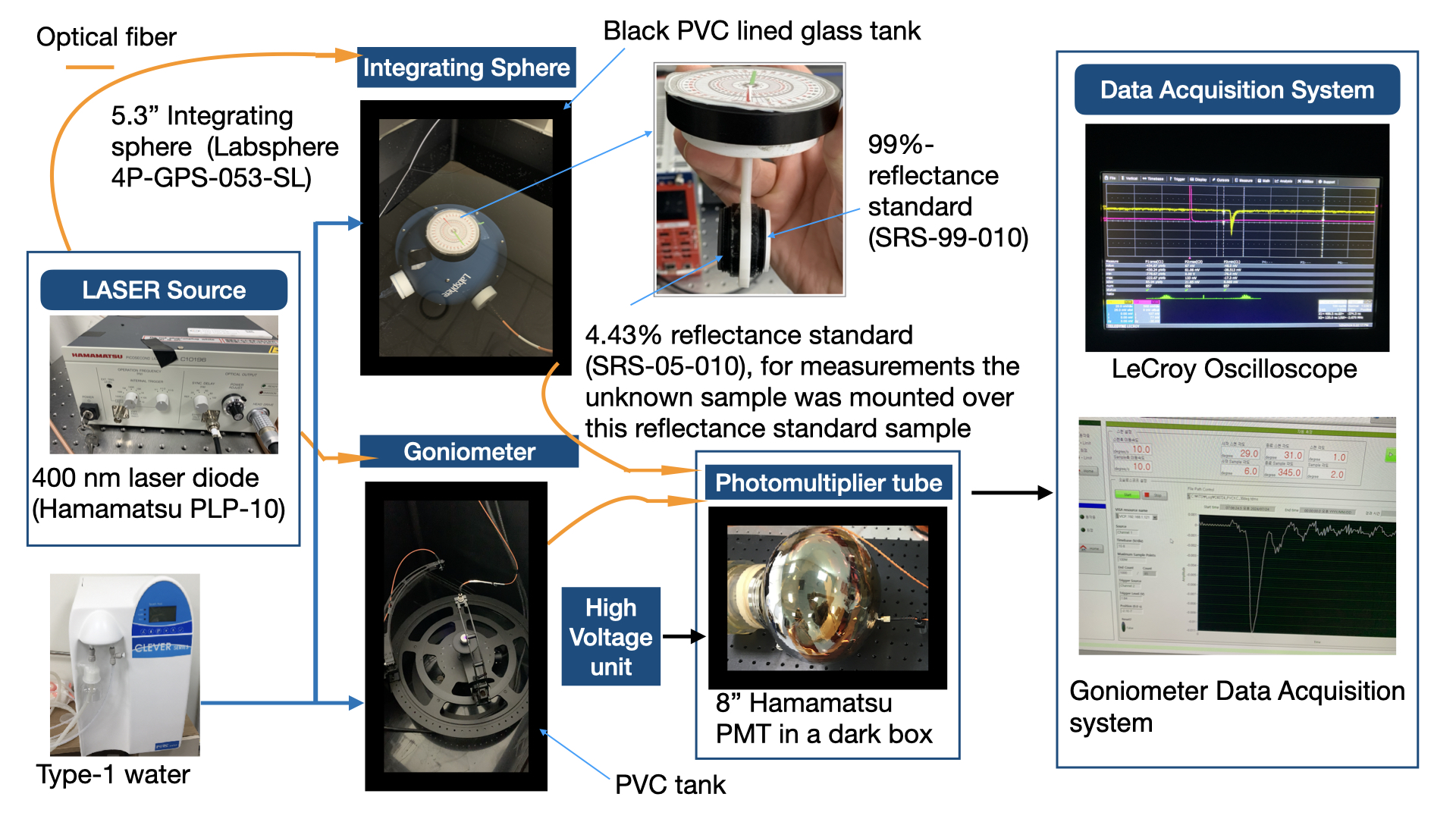}
\caption{The optical characterization setup at IBS. The laser input source (400 nm) was used alternatively {with IS and goniometer} for the total and angular measurements, {respectively}. The reflected light from the IS/goniometer was fed to the same PMT { with a 1500 V bias voltage}. For data acquisition with IS, an oscilloscope was used, however, for the goniometer data taking  the oscilloscope along with a custom-made software was employed.\label{fig:3}}
\end{figure}

We use Labsphere’s\cite{WinNT} integrating sphere (IS) as a primary tool to estimate the absolute (or total) reflectance of an optical surface in air and water. Figure~\ref{fig:2} (left) shows the integrating sphere set up for the total reflectivity measurement through a `comparison mode' \cite{Lab2}, which is simply the ratio of the mean of pulse area for the sample under consideration and the reference standard white sample with a known reflectance. For measurement of the angular profile of the reflectivity, we use a water-proof 2-dimensional goniometer developed by a South Korean company. {The right panel of Figure~\ref{fig:2}} shows the schematic for a goniometer measurement. The complete setup, with details of each module, is described in Figure~\ref{fig:3}.

\section{Challenges \& mitigation}
\label{sec:chal}
\begin{figure}[htbp]
\centering
\includegraphics[width=.35\textwidth]{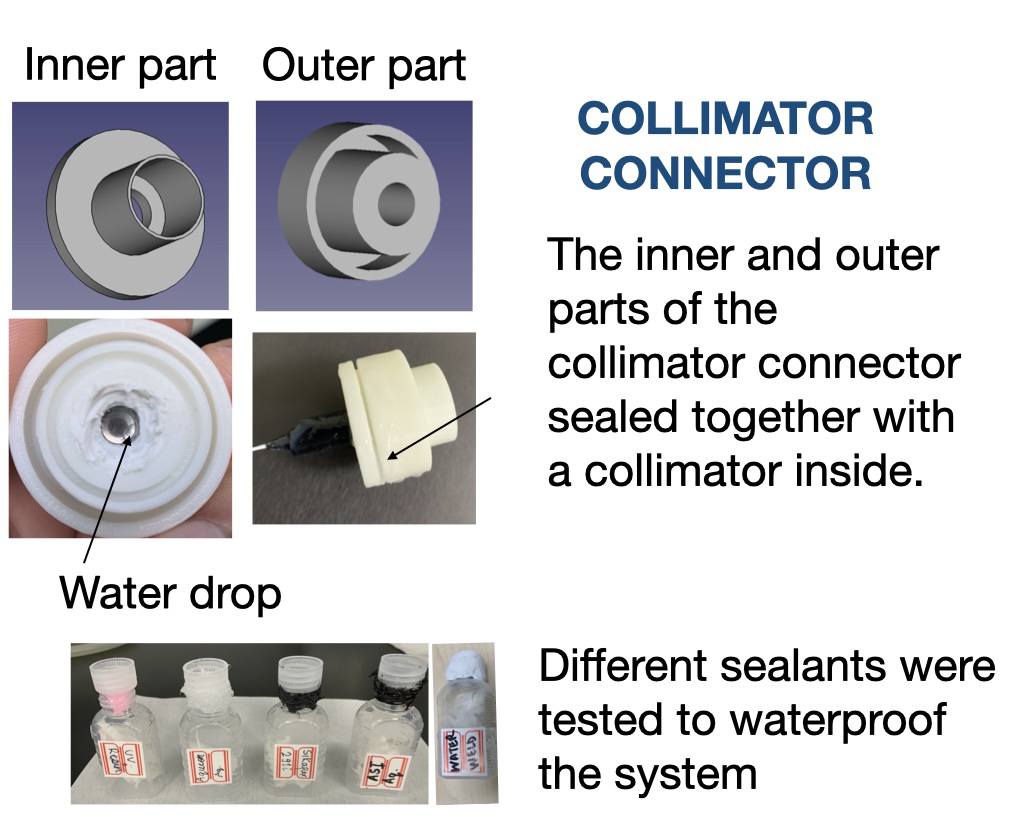}
\qquad
\includegraphics[width=.35\textwidth]{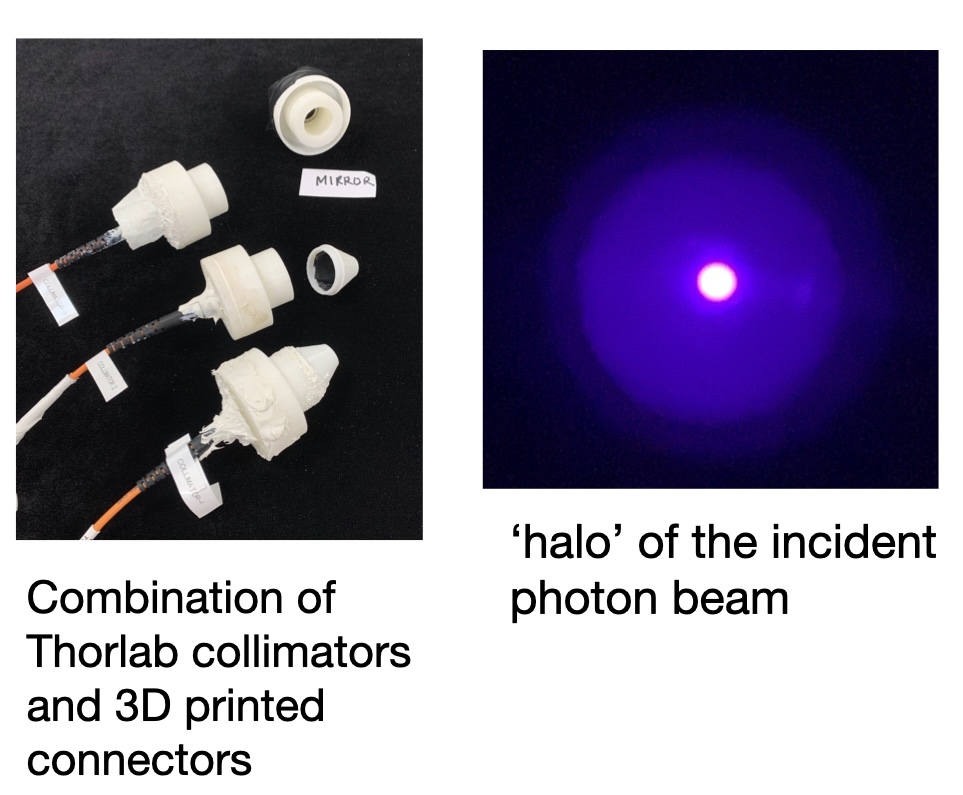}
\caption{Challenges and mitigation: Several combinations of Thorlab(\url{www.thorlabs.com}) collimators and 3D printed connectors were tried (Collimator-1 : [CFC11P-A, f= 11 mm, FC/PC ], Collimator-2: [CFC8-A, f= 7.5 mm, FC/PC], Collimator-3: [CFC5-A, f=4.6 mm, FC/PC]). \label{fig:4}} 
\end{figure}

The major challenge was designing a watertight optical coupling without losing the beam collimation substantially (see Figure~\ref{fig:4}). In that pursuit, we designed and 3D printed plastic connector and tried coupling different adjustable collimators. Further, to waterproof the collimator connectors, we tested several sealants before succeeding. A leaked connection caused droplets (fog) inside in the lens, deteriorating the measurements. {The `halo’ of the incident photon beam was another major challenge, independent of the optical medium. To mitigate this, a 3D-printed neck collar/baffle was used. These challenges contributed to the systematic uncertainties in our measurements, and efforts are ongoing to quantify them.}

\section{Results \& discussions}
\label{sec:results}
We discuss the main results in this section. Using the comparison mode scheme for the IS, as mentioned in section~\ref{sec:method}, we first evaluated the optical system performance for the diffused standard reflectance surfaces (black and white in Figure~\ref{fig:3} whose reflectance was known.  The total reflectance (in the air) of the black reference standard showed ~20 \% higher value than the known value (4.4\%), {probably due to systematic errors such as the remaining halo component}. Among the materials that we tested, results for a black PVC (sourced locally) are presented here. The overall reflectivity of the PVC was found to be around 5\% in the air and less (around 4\%) in the water. We noticed that the reflectivity values in the water are consistently lower than the corresponding values in the air{, possibly due to reduction in halo in water. However, the study of systematic uncertainties is ongoing, so the reduction in reflectance in water remains inconclusive.}

For the angular reflectance measurements using a goniometer (in the air), data taking spanned over $\theta=[0,80], \phi=[0,360]$. Scanning with $5^\circ \times 30^{\circ}$, grid size takes around 11 hours) and hence for the sake of faster data collection, a few scans with coarser angular bins were done first and specular components were located. Thereafter, for the $\theta$ and $\phi$ bins, where specular components were prominent, data taking was repeated with a finer grid. The plots for each optical surface under consideration in Figure~\ref{fig:4} were obtained by combining different data sets, keeping the primary conditions (i.e. input beam wavelength, pulse frequency, PMT bias voltage) the same. For getting smoother maps, data interpolation has been used. We can visually see the diffused reflection profile for the reference samples and a prominent specular component for the PVC. The total (summed) output intensity in each case is proportional to their reflectivity (as measured by IS). Once we have the 3D profile, for extracting the BRDF profile of the sample material and its implementation in the Geant4 simulation toolkit, we follow the prescription in ref~\cite{Nozka:11}. Fitting the 3D BRDF with multiple physics components, we can extract the coefficients for diffused, specular lobe, and specular spike. For the black PVC, using 5\% total reflectance from IS measurements, and assuming no backscatter component and moderate roughness of the surface, we get the following normalized coefficient values: 0.98 (specular lobe), 0.016 (specular spike) and negligible value for the diffused reflectivity. It is to be noted that the results shown here are preliminary and precise results will be presented in a follow-up work by the authors.
\begin{figure}[htbp]
\centering
\includegraphics[width=.45\textwidth]{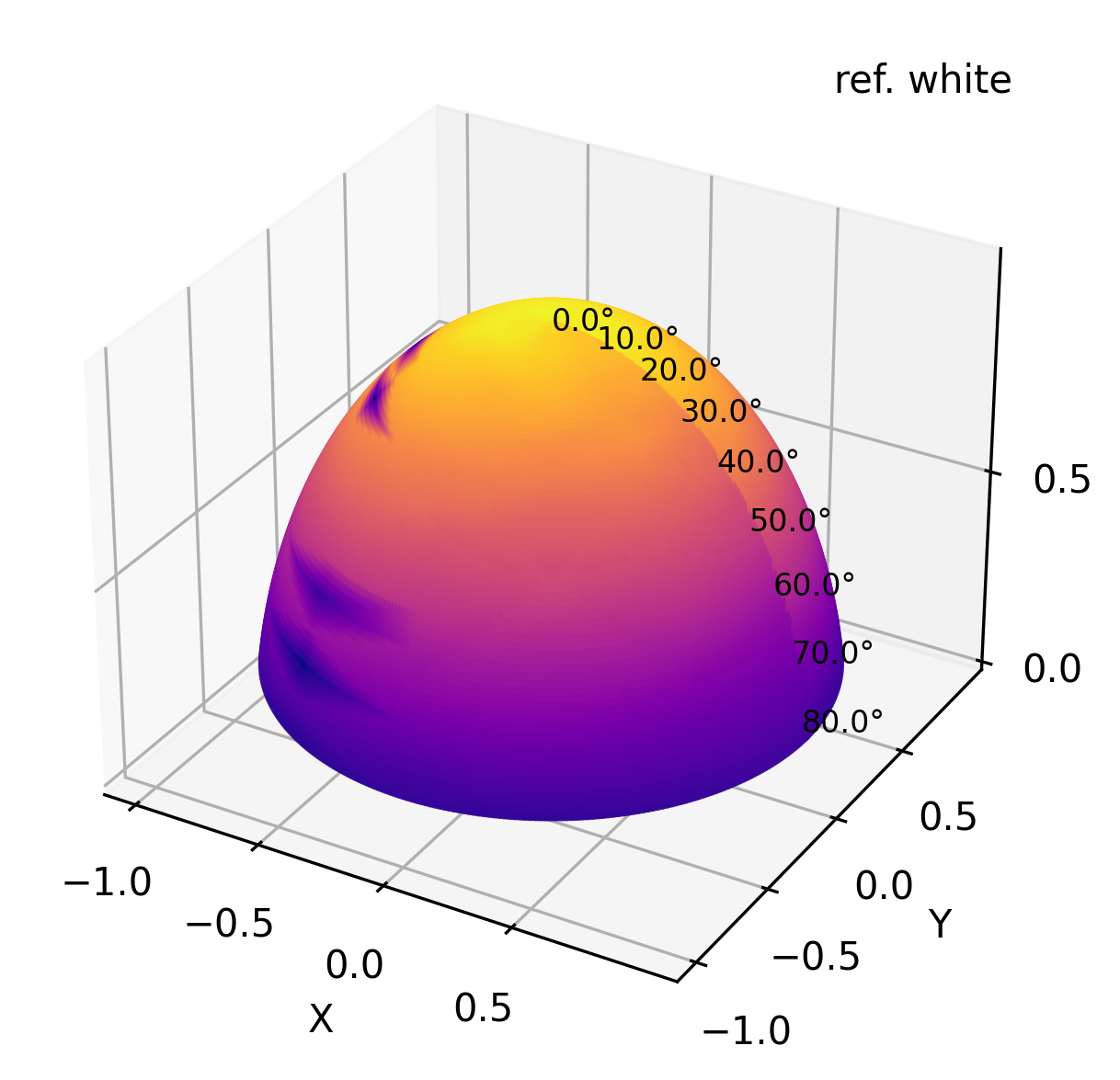}
\qquad
\includegraphics[width=.45\textwidth]{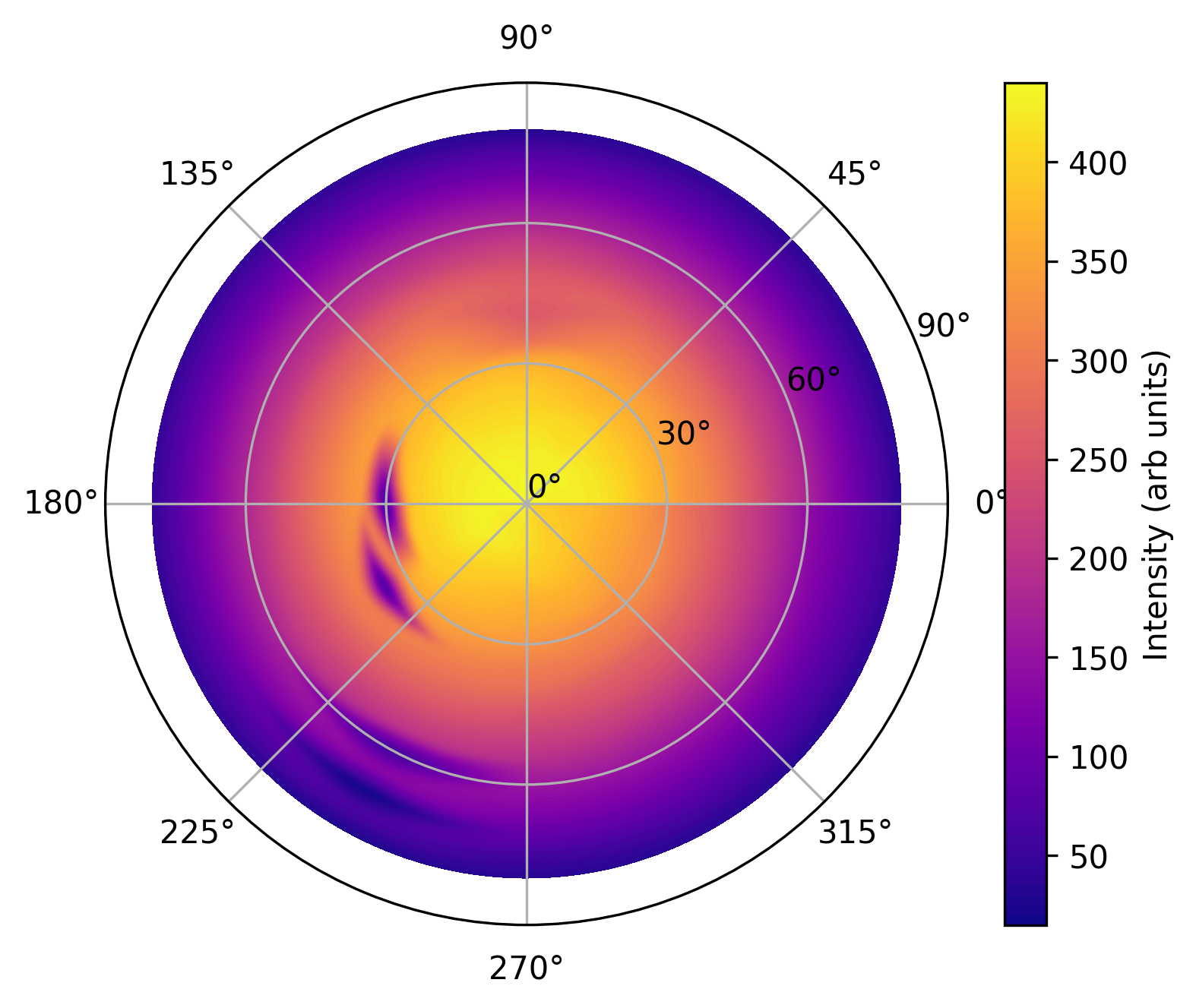}
\qquad
\includegraphics[width=.45\textwidth]{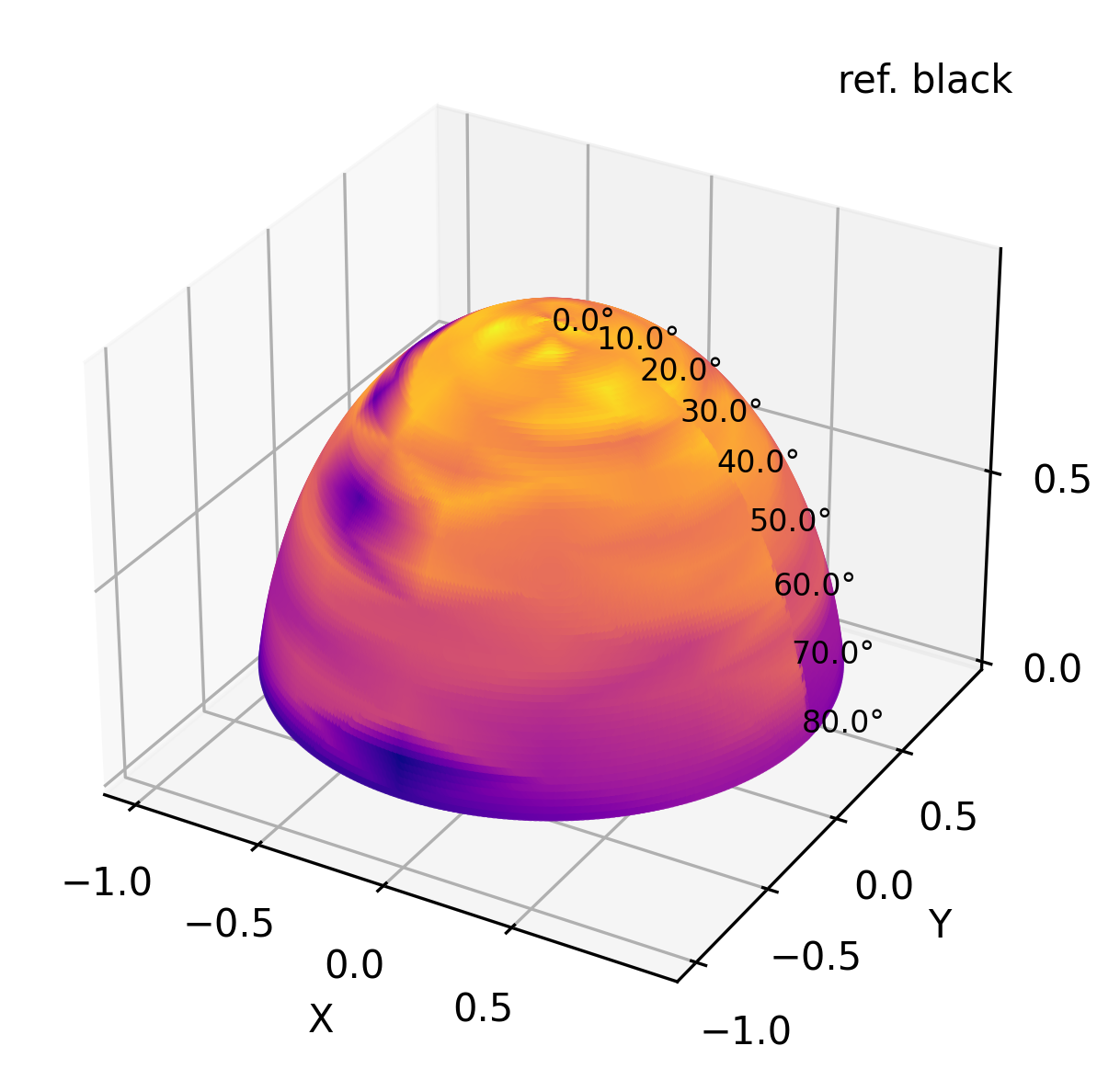}
\qquad
\includegraphics[width=.45\textwidth]{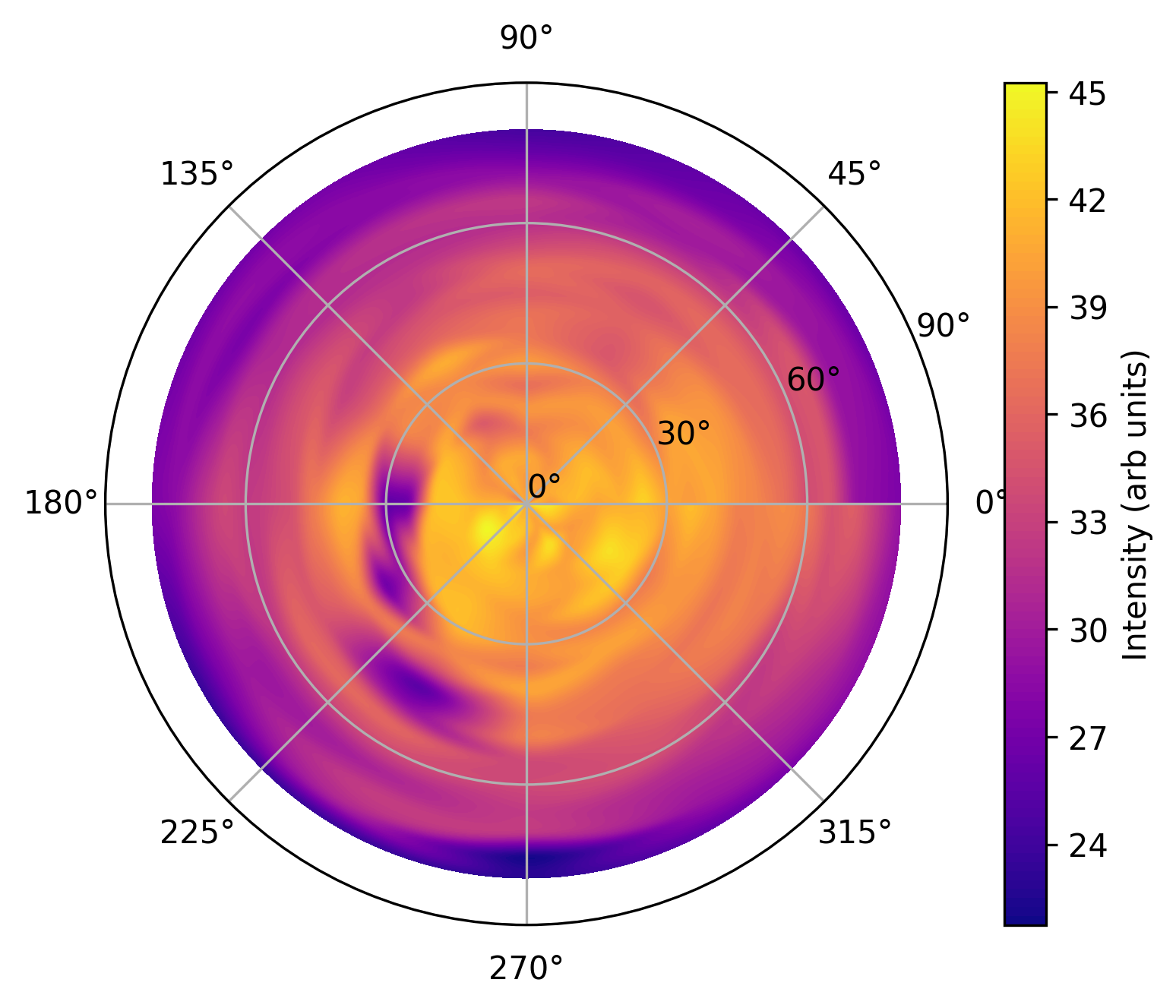}
\qquad
\includegraphics[width=.45\textwidth]{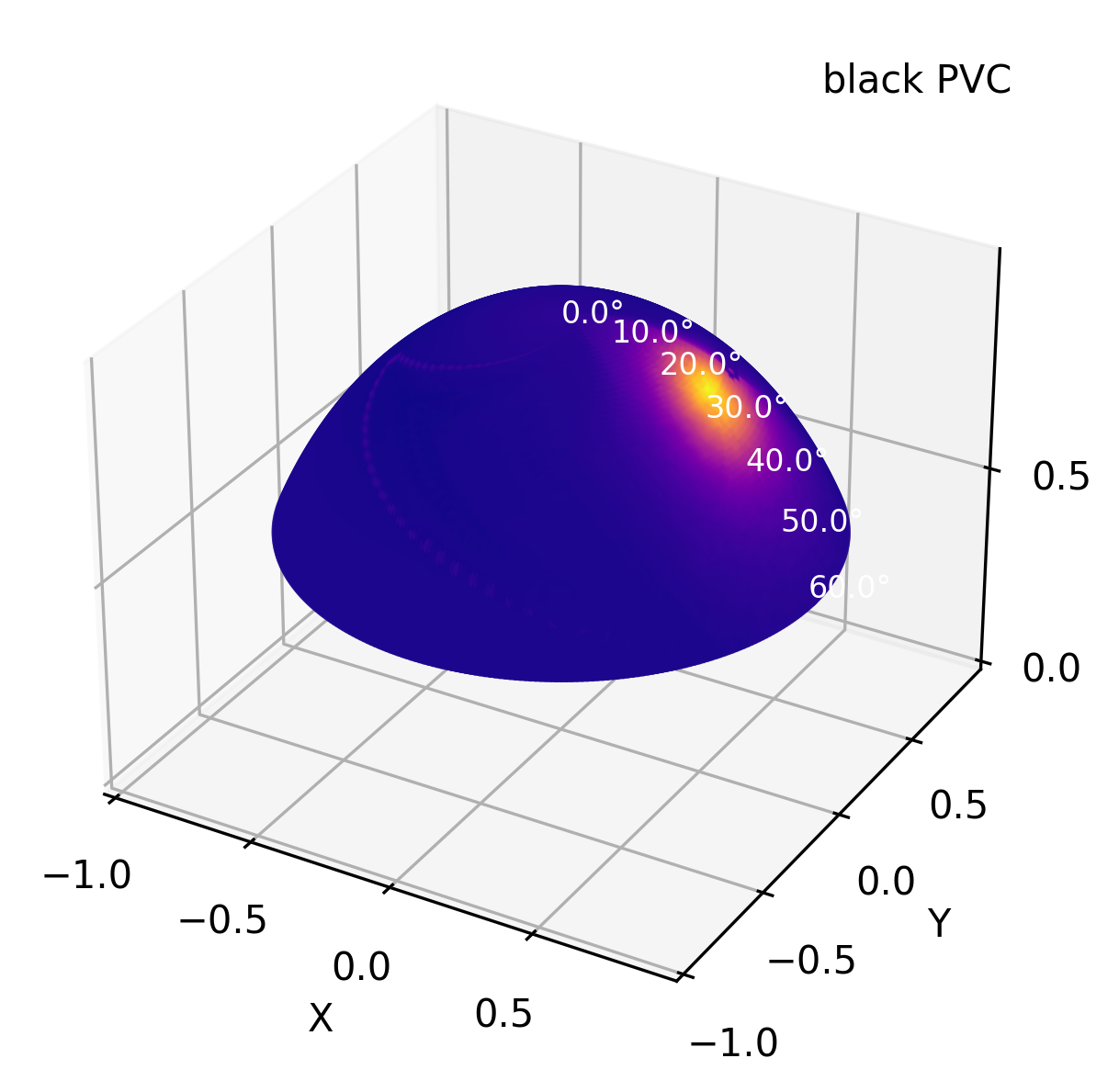}
\qquad
\includegraphics[width=.45\textwidth]{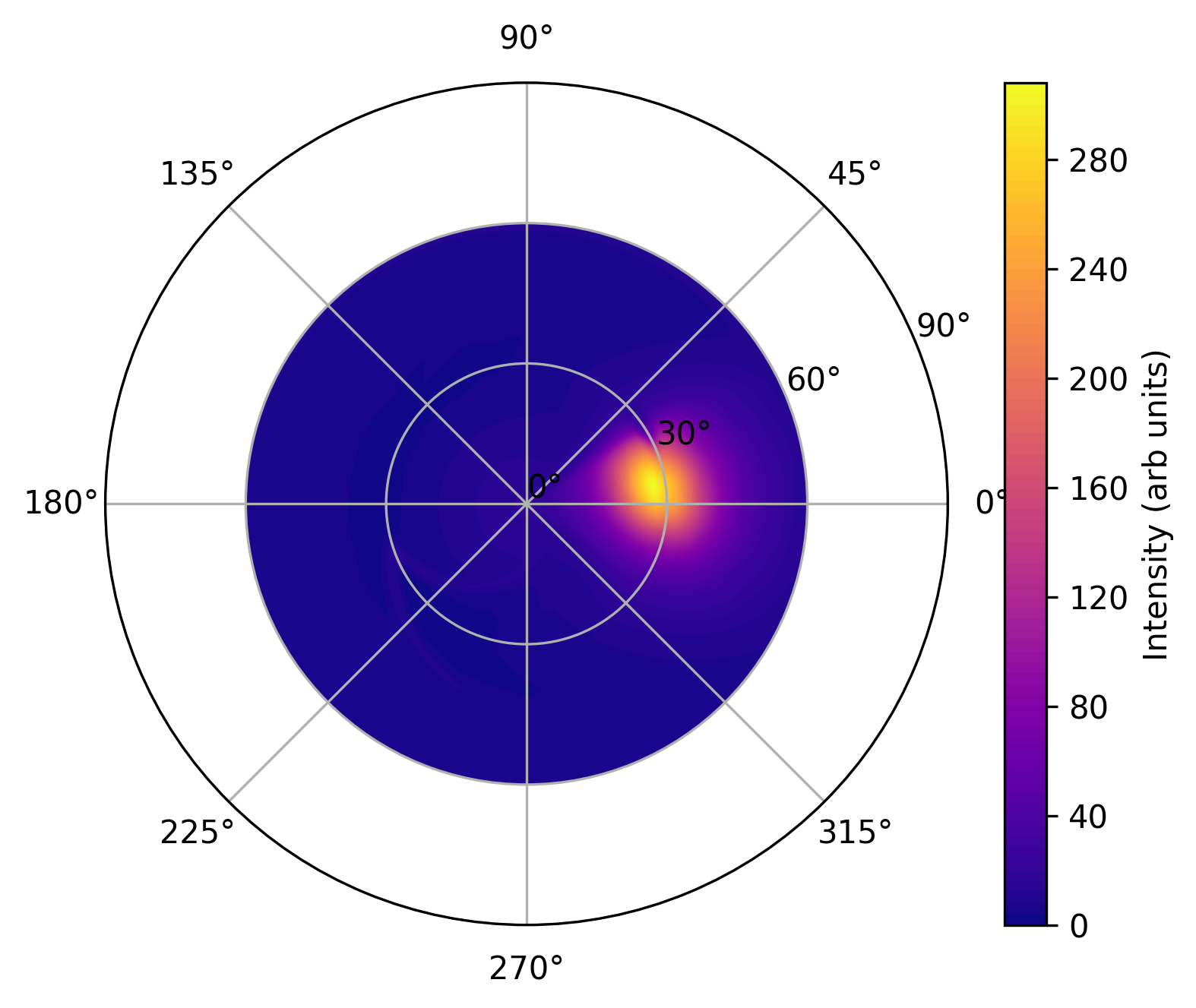}
\caption{3D surface intensity maps (left panel) showing the measurements using a goniometer (in the air) for a standard reference white (top), reference black (middle), and the black PVC (bottom) sample under consideration. The right panel shows the corresponding polar plots. $\theta_{i}=30^{\circ}$, and $\lambda=400$ nm for all the cases, and each bin represents a mean of 1000 pulses. For PVC measurements, $\theta=[0,60]$ only. For smooth visualization, the data points have been interpolated to generate this 3D map (hence the slight jumps at the azimuthal angle = 0). The low-intensity specks on these maps are due to the shadowing effect of the input arm of the goniometer.  We can visually see the diffused angular reflection for the reference samples and the specular reflection in the case of the PVC sample. The total (summed) output intensity in each case is proportional to their reflectivity as measured independently by an integrating sphere.\label{fig:5}}
\end{figure}

\section{Industrial applications \& Outlook}
\label{sec:ind}
BRDF measurements have various industrial applications where surface reflectivity, appearance, and scattering properties are critical. Some examples of BRDF applications in different industries include ensuring uniform reflectivity and anti-glare coatings for high-end screens, surface property analysis for earth observation, and analyzing tissue reflectance for medical diagnostics. Further applications comprise improving realistic lighting effects in CGI and gaming, and in the automotive industry for evaluating the appearance and glossiness of car paints.

\acknowledgments
DT gratefully acknowledges the contribution of Koun Choi, IBS, for the initial setup design and planning. This project was funded {by} the Institute of Basic Science, South Korea.  Our sincere thanks to the WCTE Collaboration and PD24 organizers.

\bibliographystyle{JHEP}{}   
\bibliography{ref}
\end{document}